\pdfoutput=1
\documentclass[conference, twocolumn, 10pt]{IEEEtran}
\usepackage{fixltx2e}
\usepackage[utf8x]{inputenc}
\usepackage[T1]{fontenc}
\usepackage{textcomp}
\usepackage[artemisia]{textgreek}
\usepackage[T1]{url}
\usepackage[final]{microtype}
\usepackage{amsmath}
\usepackage[thmmarks, amsmath]{ntheorem}
\usepackage[romanConstants, boldVectors]{scstuff}
\usepackage{units}
\usepackage{xspace}
\usepackage{bm}
\usepackage{paralist}
\usepackage{mdwtab}
\usepackage{acronym}
\usepackage{cite}
\usepackage[pdftex]{graphicx}
\usepackage[svgnames]{xcolor}
\usepackage[italian,english]{babel}
\usepackage[caption=false]{subfig}
\usepackage[unicode,pdftex]{hyperref}
\usepackage{bbding}

\hypersetup{colorlinks=true, 
  linkcolor=DarkBlue,
  citecolor=DarkBlue,
  urlcolor=DarkBlue}

\graphicspath{{./}{../Figures/}{./Figures/}}

\newcommand{\emailurl}[1]{%
  \href{mailto:#1}{\nolinkurl{#1}}}

\newacro{PWM}{Pulse Width Modulation}
\newacro{PDM}{Pulse Density Modulation}
\newacro{EMI}{Electro-Magnetic Interference}
\newacro{NTF}{Noise Transfer Function}
\newacro{STF}{Signal Transfer Function}
\newacro{SNR}{Signal to Noise Ratio}
\newacro{rms}{root mean square}
\newacro{OSR}{over-sample rate}

\newcommand{\PWM}{\ensuremath{\mathrm{PWM}}}
\newcommand{\NTF}{\ensuremath{\mathit{NTF}}}
\newcommand{\STF}{\ensuremath{\mathit{STF}}}
\newcommand{\FF}{\ensuremath{\mathit{FF}}}
\newcommand{\FB}{\ensuremath{\mathit{FB}}}

\theorembodyfont{\itshape}
\theoremheaderfont{\normalfont\bfseries}
\theoremseparator{}

\theoremstyle{nonumberplain}
\theoremheaderfont{\normalfont\bfseries}
\theorembodyfont{\normalfont}
\theoremsymbol{~~\IEEEQEDopen}

\renewcommand\normalsize{\fontsize{9.5}{11.4}\selectfont}

\begin{document}

\title{%
  Should \texorpdfstring{ΔΣ}{Delta-Sigma} Modulators Used in AC Motor Drives\\
  be Adapted to the Mechanical Load of the Motor?}

\makeatletter
\let\thanks\@IEEESAVECMDthanks
\makeatother

\author{%
  \IEEEauthorblockN{Sergio Callegari}
  \IEEEauthorblockA{University of Bologna,\\
    ARCES/DEIS, Italy\\
    \url{sergio.callegari@unibo.it}} \and \IEEEauthorblockN{Federico Bizzarri}
  \IEEEauthorblockA{Polytechnic of Milan\\
    DEI, Italy\\
    \url{bizzarri@elet.polimi.it}}%
  \thanks{This is a post-print version of a paper published as in the
    Proceedings of 2012 IEEE International Conference on Electronics, Circuits,
    and Systems (ICECS), pp.\@ 849-852, Dec.\@ 2012. Available through DOI
    \href{http://dx.doi.org/10.1109/ICECS.2012.6463619}%
    {10.1109/ICECS.2012.6463619}. To cite this document please use the
    published version data.
    \protect\\[1ex]
    Copyright © 2012 IEEE. Personal use of this material is permitted. However,
    permission to use this material for any other purposes must be obtained
    from the IEEE by sending a request to
    \url{pubs-permissions@ieee.org}.\protect\\[-2ex]}}

\PrerenderUnicode{©}

\initializeplaintitle[%
  \def\texorpdfstring#1#2{#2}]
\hypersetup{%
  pdftitle=\plaintitle),
  pdfauthor={Sergio Callegari, Federico Bizzarri}
}

\maketitle

\begin{abstract}
  We consider the use of ΔΣ modulators in ac motor drives, focusing on the many
  additional degrees of freedom that this option offers over
  \ac{PWM}. Following some recent results, we show that it is possible to fully
  adapt the ΔΣ modulator \ac{NTF} to the rest of the drive chain and that the
  approach can be pushed even to a fine adaptation of the \ac{NTF} to the
  specific motor loading condition. We investigate whether and to what extent
  the adaptation should be pursued. Using a representative test case and
  extensive simulation, we conclude that a mild adaptation can be beneficial,
  leading to \ac{SNR} improvements in the order a few \unit{dB}, while the
  advantage pushing the adaptation to the load tracking is likely to be
  minimal.
\end{abstract}

\acresetall
\acused{SNR}
\acused{rms}

\section{Introduction}
\label{sec:intro}
Switched-mode power conversion keeps gaining momentum due to its efficiency and
flexibility. An important application is ac drives where induction motors are
fed by inverters so that both frequency and voltage can be finely varied. Two
major approaches exist, relying on \ac{PWM} or \ac{PDM}.

In \ac{PWM}, frequency and voltage control is achieved by varying the duty
ratio of the inverter switches \cite{Holmes:PWMPC-2003}. To this aim, a fixed
frame frequency $f_{\PWM}$ is established, so that the time axis is divided in
frame intervals. For each interval, the modulator produces a pulse, selecting
its width as needed. Conversely, in \ac{PDM} voltage and frequency are
controlled by varying the concentration of pulses \cite{Jacob:EPA-5-7}. There
is no frame concept, but a thin pulse duration $T$ is established, so that the
time axis is divided in pulse intervals. As time flows, for each pulse interval
the modulator can decide whether to generate a pulse or not, thus varying the
output density as needed.

While effective, \ac{PWM} is sometimes criticized for its fixed frame frequency
that may lead to evident harmonic clusters \cite{Boys:EPA_B-137-4} in voltage
and current spectra. This may cause \ac{EMI} or acoustic noise
\cite{Lowery:TIA-30-2}. Randomized schemes, where either the pulse width,
position or frame frequency are perturbed may reduce this effect
\cite{Stankovic:I3P-90-5, Balestra:TIEICEE-E87C-1, Callegari:EL-38-12}. Yet,
they may lead to other issues and certainly increase system complexity.

This is one of the reasons why \ac{PDM}, is now actively investigated as a
\ac{PWM} substitute \cite{Jacob:EPA-5-7}. Lacking a frame structure, it can be
inherently more compliant to \ac{EMI} regulations. Furthermore, the higher
apparent complication of \ac{PDM} is often a myth. First of all, very practical
implementations now exist, typically relying on ΔΣ modulation
\cite{Schreier:UDSDC-2004}, at times practiced with specialized quantizers
(e.g., hexagonal \cite{Luckjiff:TCAS1-50-8} or vector type
\cite{Jacob:EPA-5-7}). Secondly, the perception of complication mostly arises
from the much wider set of tuning options that it offers. Operating in a wider
design space is obviously harder. This is evident just by looking at the number
of \emph{adjustable knobs}. In \ac{PWM} only the frame frequency and resolution
need to be set. In ΔΣ modulation, which is already a special, restricted kind
of \ac{PDM}, the full set of coefficients for the filters inside the modulator
is to be chosen by the designer, in addition to the sample rate.

In this paper we deal with this extra flexibility and its
exploitation. Particularly, we try to answer a specific question.  For a ΔΣ
modulator used to drive an ac motor, there is an obvious desire to set its
parameters according to the specifications of the power bridge, the motor
itself and any other elements sitting between the two (e.g., a passive filter)
i.e, to \emph{adapt} the modulator to the rest of the drive.  Now, given that
the motor behavior can vary, even significantly, with the mechanical load
applied to its shaft, should one try to track these changes, making the
modulator parameters vary accordingly? Or would this be a useless attempt at
over optimize the system, with negligible benefit in terms of measured
performance?

To answer this question we simulate a realistic motor, modeled at different
loading conditions and with bridge commands synthesized by different ΔΣ
modulators, some of which explicitly designed after the motor behavior at a
specific mechanical loading. For the modulator optimization we exploit a recent
result that enables its adaptation to the user of the ΔΣ streams
\cite{Callegari:TCAS1-Submitted-2012,
  Callegari:DeltaSigmaRep-2012}. Eventually, for each loading condition, we
compare the optimal behavior with that obtained from the modulator optimized at
another operating condition.

\section{Modeling}
\begin{figure}[t]
  \begin{center}
    \includegraphics[width=\lw]{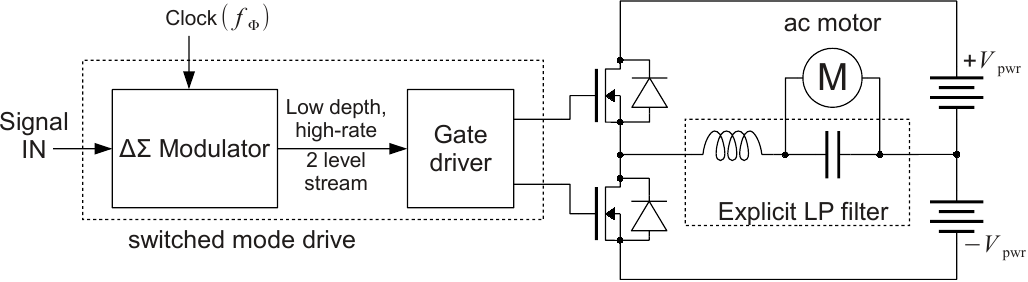}
  \end{center}
  \vspace{-1ex}
  \caption{Idealized, single phase simulation setup.}
  \label{fig:block-dia}
\end{figure}

In our analysis, we exploit a realistic motor model. Yet, we consider an
idealized setup restricted to a single phase feed-forward control as in
Fig.~\ref{fig:block-dia}. This is to decouple the matter under exam from any
other second order effect, such as those possibly arising from
multiphase/multilevel quantization or feed-back. We also practice a direct
connection between the power bridge and the motor (i.e, we remove the filter
from the architecture in Fig.~\ref{fig:block-dia}), leaving all the signal
smoothing to the reactive effects provided by the motor itself. While this is
unusual for large drives, it lets the consequences of impedance variations
associated to changes in the mechanical load be more evident, avoiding their
hiding behind other large, fixed reactive effects.

For what concerns the electric machine, we start from the usual equations for a
3-phase induction motor. Under the basic hypothesis of writing the equations:
(i) in a stationary reference frame; (ii) assuming null rotor voltages; and
(iii) taking the equivalent $dq$-axis representation, the following
differential system is derived \cite{Fitzgerald:EM-1990}
{\fontsize{9}{10.8}%
  \begin{align}
    \frac{d i_{sd}}{dt} &= \frac{-R_s i_{sd} + \frac{ \omega_m P L_m^2}{2
        L_{r}} i_{sq} + \frac{R_r L_m}{L_{r}} i_{rd}  + \frac{\omega_m P
        L_m}{2} i_{rq} + v_{sd}}{\Delta\cdot L_{s} }\notag\\ 
    \frac{d i_{sq}}{d t} &= \frac{- \frac{\omega_m P L_m^2}{2 L_{r}} i_{sd} -R_s
      i_{sq} -\frac{\omega_m P L_m}{2} i_{rd} + \frac{R_r  L_m}{L_{r}} i_{rq}
      + v_{sq}}{\Delta\cdot L_{s} }\notag\\ 
    \frac{d i_{rd}}{d t} &= \frac{\frac{R_s  L_m}{L_{s}} i_{sd} -\frac{\omega_m
        P L_m}{2} i_{sq} -R_r i_{rd} -\frac{\omega_m P L_r}{2} i_{rq} -
      \frac{L_m}{L_{s}} v_{sd}}{\Delta\cdot L_{r} }\notag\\ 
    \frac{d i_{rq}}{d t} &= \frac{\frac{\omega_m P L_m}{2} i_{sd} + \frac{R_s
        L_m}{L_{s}} i_{sq} + \frac{\omega_m P L_{r}}{2} i_{rd} -R_r i_{rq} -
      \frac{L_m}{L_{s}} v_{sq}}{\Delta\cdot L_{r} }\notag\\ 
    \frac{d \omega_m}{dt} &= \frac{\frac{3}{4}(i_{sq} i_{rd} - i_{sd}
      i_{rq}) P L_m - \Lambda(t) - B \omega_m}{J}.
    \label{eq:imode2}
  \end{align}%
}%
where: $\Delta = 1- \nicefrac{L_m^2}{(L_{s} L_{r})}$, $i_{\alpha,
  \beta}$ and $v_{\alpha, \beta}$ are the currents and voltages referred to the
motor stator or rotor (if $\alpha=s$ or $\alpha=r$, respectively), once
projected on the $\beta$-axis for $\beta \in \{d,q\}$; $\omega_m$ is the rotor
mechanical angular speed; $\Lambda(t)$ is a time-varying load. Other parameters
and symbols are explained in Tbl.~\ref{tbl:motor-params} that also reports the
values used for simulation (which are the same used in
\cite{Bizzarri:ISCAS-2012}, for the sake of an easy confrontation).

\begin{table}[b]
  \caption{ac motor parameters and values used in simulation}
  \label{tbl:motor-params}
  \vspace{-3ex}
  \begin{center}
    \begin{tabular}{|Mc|Mc|m{0.55\lw}|}
      \hlx{hv}
      P & 4 &
      Number of motor poles\\
      B & \unit[25\cdot 10^{-3}]{N\,m\,s} & 
      Damping constant (dissipation due to windage and friction)\\
      J & \unit[25\cdot 10^{-3}]{kg\,m^2} &
      Moment of inertia\\
      R_s & $\unit[17.7]{\Ohm}$ &
      Stator dissipative effects\\
      R_r & $\unit[13.8]{\Ohm}$ &
      Rotor dissipative effects\\
      L_s & \unit[459.2 \cdot 10^{-3}]{H} &
      Total 3-phase stator inductance\\
      L_r & \unit[457.0\cdot 10^{-3}]{H} &
      Total 3-phase rotor inductance\\
      L_m & \unit[442.5\cdot 10^{-3}]{H} &
      Magnetizing inductance\\
      f_w & \unit[50]{Hz} &
      Nominal driving voltage angular frequency\\
      V_w & \unit[320]{V} &
      Nominal driving voltage peak amplitude\\
      \hlx{h}
    \end{tabular}
  \end{center}
\end{table}

\begin{figure}[t]
  \begin{center}
    \includegraphics[width=\lw]{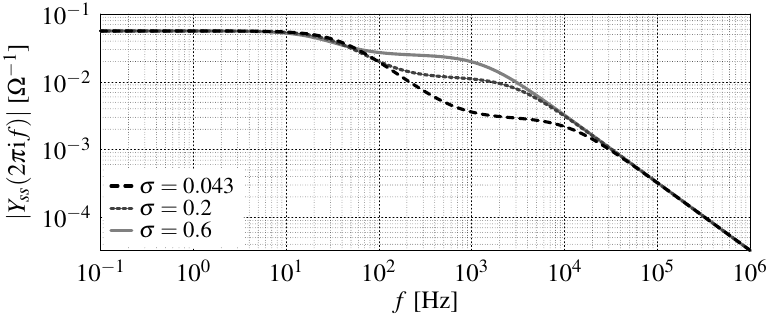}
  \end{center}
  \caption{Transfer function due to the motor admittance at different slip
    values.}
  \label{fig:TFs}
\end{figure}

Evidently: (i) the model is non-linear; and (ii) it depends on the load
$\Lambda$. For what concerns the first point, it is worth observing that if
$\omega_m$ has an almost constant value, one can approximate away the
non-linearity, obtaining the well known \emph{linear} model describing the
steady-state dynamics of the motor working at the $\omega_e= \nicefrac{\omega_m
  P}{2}$ rotor constant electrical angular frequency.  This is related to the
angular frequency $\omega_w=2\pi f_w$ of the three-phase sinusoidal drive
voltage (with amplitude $V_w$) through the slip coefficient
$\sigma=1-\nicefrac{\omega_e}{\omega_w}$ running from $0$ (no-load condition)
to $1$ (blocked rotor). From this linear model, one can eventually compute the
admittances from each of the three phase stator voltages to the corresponding
phase stator current, namely
\begin{equation}
  Y_{ss}(\ii \omega)=\frac{\frac{R_r}{L_s L_r} + \ii \frac{\sigma}{L_s} \omega}
  { \frac{R_r R_s}{L_r L_s} +\ii(\frac{R_r}{L_r}+ \sigma \frac{R_s}{L_s})\omega
    -\sigma \Delta \omega^2} .
  \label{eq:TF}
\end{equation}
Interestingly, in this linearized model, the mechanical loading effects seem to
disappear. As a matter of fact, they do not. They are only hidden inside the
slip parameter. A higher load requires a higher motor torque that can only be
obtained by augmenting the motor currents or accepting a higher
slip. Fig.~\ref{fig:TFs} shows the frequency response associated to $Y(\ii 2\pi
f)$ for three different slip values (self-friction, $\sigma=0.043$, nominal
slip $\sigma=0.2$ and large slip $\sigma=0.6$) and for the motor parameters in
Tbl.~\ref{tbl:motor-params}. Evidently, the change is quite significant. It is
worth underlining that this behavior, with a pole-zero couple that moves as
$\sigma$ is increased, is not limited to the example setup, but rather typical.

The linearized model~\eqref{eq:TF} enables a rapid evaluation of some aspects
of the switched drive performance. Specifically, it lets one foresee the
deviation of the motor windings currents from ideal sine waves. Such deviation
results in high frequency components and on \ac{EMI}. Furthermore, since
currents are ultimately responsible for the motor torque, it may cause torque
fluctuation, vibration and noise. Consequently, conformance of the winding
currents to expected sinusoidal profiles, quantified through \ac{SNR}, can be
taken as an important quality figure.  Formally, \ac{SNR} is defined as
$\nicefrac{I^2_N}{I^2_S}$, where $I_S$ is the effective value of the ideal
drive and $I_N$ is the \ac{rms} current value associated to the switched
actuation artifacts.  One has
\begin{equation}
  I^2_N=\int_{0}^{\infty} \Psi^2_N(\ii 2\pi f) \abs{Y(\ii 2\pi f)}^2 df
  \label{eq:int}
\end{equation}
where $\Psi^2(\ii 2\pi f)$ is the spectral density associated to the actuation
artifacts on the voltage drive. Obviously, in Eqn.~\eqref{eq:int} there may be
an additional multiplicative term $\abs{G(\ii 2\pi f)}^2$ whenever a filter
$G(s)$ is used to feed the motor, but we have chosen not to implement it. From
Eqn.~\eqref{eq:int}, the importance of $Y(\ii 2\pi f)$ is self evident, given
that $\Psi^2(\ii 2\pi f)$ depends and can be determined from the drive.

\section{ΔΣ modulator in the drive} 
\label{sec:shaping}
\begin{figure}[t]
  \begin{center}
    \includegraphics[scale=0.65]{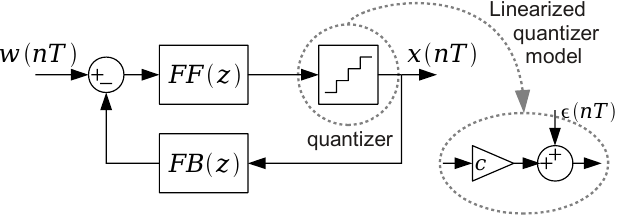}\\
  \end{center}
  \caption{ΔΣ modulator and its approximated linear model.}
  \label{fig:ds}
\end{figure}

One of the advantages of using a ΔΣ modulator in the drive is to exploit its
noise shaping abilities adapting them to the signals to be actuated and
possibly to the drive chain. Recall that a ΔΣ modulator can be modeled by the
standard architecture in Fig.~\ref{fig:ds}, where $T$ is the sample time. For
an approximated analysis, the model gets linearized as illustrated in the
circle. In this case, one can identify a \ac{NTF}, $\NTF(z)$ from the
quantization noise $\epsilon(nT)$ to the output and a \ac{STF}, $\STF(z)$ from
the modulator input $w(nT)$ to the output. Once the \ac{STF} is taken to be
unitary as it is normally the case, the modulator filters $\FF(z)$ and $\FB(z)$
can be designed to obtain a desired \ac{NTF} as long as some constraints on the
\ac{NTF} itself are respected \cite{Schreier:UDSDC-2004}.

This is particularly interesting with reference to Eqn.~\eqref{eq:int}. In
fact, for a correctly operating modulator, $\epsilon(nT)$ can be assumed to be
approximately uncorrelated to the input, white, and uniformly distributed
within $[\nicefrac{-\Delta}{2}, \nicefrac{\Delta}{2}]$ where $\Delta$ is the
quantization step \cite{Schreier:UDSDC-2004}. Consequently, $\Psi^2(\ii 2\pi
f)$ in~\eqref{eq:int} happens to be known and approximately equal to
$\frac{\Delta^2}{12} \sinc^2(f T)\abs{\NTF\left(\ee^{\ii 2\pi
      \frac{f}{f_s}}\right)}^2$, where the usual $z \leftrightarrow \ee^{\ii
  2\pi \frac{f}{f_s}}$ substitution is adopted, $f_s=\nicefrac{1}{T}$, a
zero-order hold is assumed at the modulator output, and $\sinc$ indicates the
normalized function $\nicefrac{\sin(\pi x)}{(\pi x)}$. Hence, $I_N^2$ is
\begin{equation}
  I^2_N=\frac{\Delta^2}{12}\int_{0}^{\infty}
  \sinc^2(f T)
  \abs{\NTF\left(\ee^{\ii 2\pi \frac{f}{f_\Phi}}\right)}^2 
  \abs{Y(\ii 2\pi f)}^2 df .
  \label{eq:int2}  
\end{equation}

From Eqn.~\eqref{eq:int2}, the opportunity of optimizing $\NTF(z)$ adapting it
to the motor transfer function $Y(s)$ is quite evident. Thus, the following
questions arise: (i) is it worth selecting a \emph{custom} \ac{NTF} rather than
using a conventional design?  (ii) does it make sense to continuously modify
the \ac{NTF} to track the variations of $Y(\ii 2\pi f)$ due to slip changes?

To ease the answer, one may first observe that thanks to the low pass nature of
$Y(\ii 2\pi f)$, and as long as the modulator sample frequency $f_s$ is
sufficiently high, the integral in Eqn.~\eqref{eq:int2} can safely be
approximated as
\begin{equation}
  I^2_N=\frac{\Delta^2}{12\pi}\int_{0}^{\pi}
  \abs{\NTF\left(\ee^{\ii \omega}\right)}^2 
  \abs{\hat Y\left(\ee^{\ii \omega}\right)}^2 df .
  \label{eq:int3}    
\end{equation}
where $\hat Y(z)$ is a discrete time version of $Y(s)$. Then, it is worth
noting, that due to the specific constraints posed by the modulator
architecture, the integral cannot be minimized in \emph{naive} ways by
nullifying the \ac{NTF} or by making it capable of concentrating \emph{all} the
quantization noise at the single frequency where $\hat Y\left(\ee^{\ii
    \omega}\right)$ attenuates most. Conversely, as explained in
\cite{Schreier:UDSDC-2004}, one needs to assure that $\abs{\NTF\left(\ee^{\ii
      \omega}\right)}$ never exceeds certain values, is non-negligible in
sufficiently large intervals and is realized by some $\NTF(z)$ whose impulse
response has a unitary first coefficient
\cite{Schreier:UDSDC-2004}. Intuitively, in presence of such constraints, the
integral can be reduced by taking $\abs{\NTF\left(\ee^{\ii\omega}\right)}$ to
be approximately the \emph{inverse} of $\abs{\hat Y\left(z\right)}$, as
discussed in \cite{Callegari:TSP-58-12}. This may require some manual
adjustment to obtain a stable $\NTF(z)$. Then, further adjustment may be
required on gain and range (namely the difference between maximum and minimum
gain values) to satisfy the modulator requirements. Alternatively, one can
choose the method in \cite{Callegari:TCAS1-Submitted-2012,
  Callegari:DeltaSigmaRep-2012} to directly obtain a suitable $\NTF(z)$ through
a better formalized path.

Following one of the two approaches above, one may obtain different \acp{NTF}
matched to specific slip values. These can be checked against a conventional
\ac{NTF} using expression~\eqref{eq:int3} to evaluate the convenience of custom
\acp{NTF} over conventional ones. Furthermore, one can evaluate some custom
\acp{NTF} designed for different slip values one against the other to see the
convenience of tracking slip changes by a continuous \ac{NTF} adjustment.

\section{Experimental results}
In our experiments, we refer to the parameter set in
Tbl.~\ref{tbl:motor-params}. Furthermore, we assume that the actuation
frequency is comprised in a \unit[[0-50]]{Hz} interval. Finally, for the ΔΣ
modulator we take an \ac{OSR} equal to 1000, setting the sample frequency at
\unit[100]{kHz}.

\begin{figure}[t]
  \begin{center}
    \includegraphics[width=\lw]{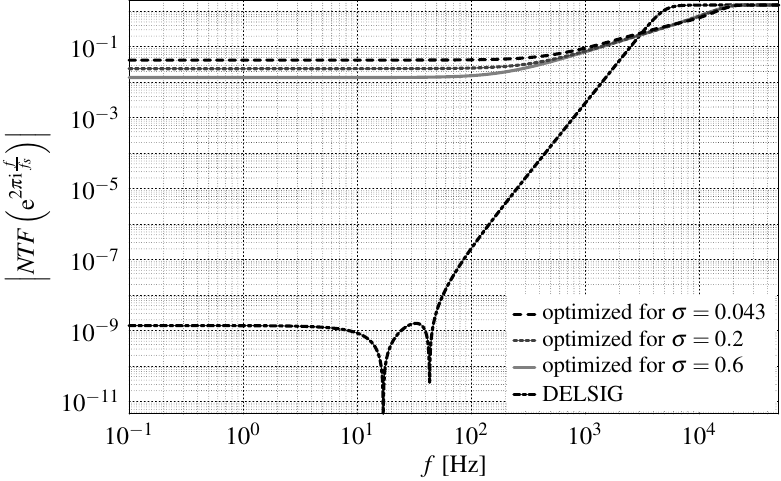}
  \end{center}
  \caption{Conventional (DELSIG) \ac{NTF} and \acp{NTF} optimized at specific
    motor loads and slips.}
  \label{fig:ntfs}
\end{figure}

With this, the diagram in Fig.~\ref{fig:ntfs} shows a standard \ac{NTF}
obtained for a 4\Us{th} order modulator through the popular design assistant
DELSIG \cite{Schreier:DELSIG}, together with three \acp{NTF} optimized for the
same slip values used for Fig.~\ref{fig:TFs}. These are 8\Us{th} order and
obtained following \cite{Callegari:TCAS1-Submitted-2012,
  Callegari:DeltaSigmaRep-2012}. The comparison of \acp{NTF} corresponding to
different modulator orders should not appear unfair, since: (i) conventional
methodologies cannot safely go beyond order 4 for this setup; and (ii) the
difference between the \acp{NTF} in the two diagram is mostly due to the
different design methodology rather than the different order. Indeed, the
striking contrast between the diagrams owes to the fact that conventional
methodologies are only based on the \ac{OSR}, while the other \acp{NTF} also
explicitly take into account the need to reduce the value of
expression~\eqref{eq:int3}.

The effectiveness of the various \acp{NTF} is reported in
Tbl.~\ref{tbl:perf-freq} which shows the \ac{SNR} obtained from
Eqn.~\eqref{eq:int3}, considering the motor model at the various slip
conditions. Data has been obtained selecting a \unit[50]{Hz} operation at
\unit[60]{\%} of the motor nominal voltage (namely \unit[190]{V} peak).

\begin{table}[t]
  \caption{\ac{SNR} in \unit{dB} obtained using the various \acp{NTF} in
    combination with the motor at different mechanical load and slip
    conditions. Data obtained from Eqn.~\eqref{eq:int3}.}
  \begin{center}
    \begin{tabular}{|c|c|c|c|}
      \multicolumn{1}{c|}{} & 
      \multicolumn{3}{c|}{Mechanical Loading Condition}\\
      \multicolumn{1}{c|}{} &
      $\sigma=0.043$ & $\sigma=0.2$ & $\sigma=0.6$\\
      \hlx{hv}
      Standard NTF & 24.72 \XSolid & 29.48 \XSolid & 35.13 \XSolid\\
      \hlx{hv}
      NTF optimized for $\sigma=0.043$ & 29.60 \Checkmark & 33.98 & 39.39\\
      \hlx{hv}
      NTF optimized for $\sigma=0.2$ & 29.50 & 34.11 \Checkmark & 39.64 \\
      \hlx{hv}
      NTF optimized for $\sigma=0.6$ & 29.45 & 34.09 & 39.67 \Checkmark\\
      \hlx{h}
    \end{tabular}
  \end{center}
  \label{tbl:perf-freq}
\end{table}

From the tabled data, a well perceivable advantage, quantifiable in
approximately \unit[3-5]{dB} (i.e. an approximate halving of the noise), seems
to be achievable by using an \ac{NTF} designed in accordance to the motor model
(rows 2 to 4), with respect to a standard \ac{NTF} designed considering the
\ac{OSR} only (row 1). Furthermore, considering rows 2 to 4, note that in each
column the best value can always be found on the diagonal. In other words, and
not surprisingly, the best performance at any slip condition appears to be
obtainable using the \ac{NTF} designed according to the motor admittance at
that particular slip condition. Nonetheless, the advantage that the best setup
offers over the others looks minimal. This indicates that optimizing the
\ac{NTF} for a specific slip may not be truly convenient. Indeed, such a
conclusion could already be expected from the curves in Fig.~\ref{fig:ntfs},
where the optimized \ac{NTF} does not change much by changing the slip level
for which the optimization is done.

To be extremely scrupulous, one may want not to limit the analysis to data
obtained by Eqn.~\eqref{eq:int3}, due to the many approximations practiced to
achieve such model. These basically amount to the linearization of both the
motor and the modulator models. Among the two, the second one is certainly the
most critical. Thus, it is worth repeating the test by means of a time domain
simulation of the modulator, based on a realistic quantizer.
\begin{table}[t]
  \caption{\ac{SNR} in \unit{dB} obtained using the various \acp{NTF} in
    combination with the motor at different mechanical load and slip
    conditions. Data obtained from time domain simulations with a nonlinear
    modulator model and a linearized motor model.}
  \begin{center}
    \begin{tabular}{|c|c|c|c|}
      \multicolumn{1}{c|}{} & 
      \multicolumn{3}{c|}{Mechanical Loading Condition}\\
      \multicolumn{1}{c|}{} &
      $\sigma=0.043$ & $\sigma=0.2$ & $\sigma=0.6$\\
      \hlx{hv}
      Standard NTF & 25.87 \XSolid & 30.73 \XSolid & 36.39 \\
      \hlx{hv}
      NTF optimized for $\sigma=0.043$ & 28.04 & 31.97 & 35.39 \XSolid \\
      \hlx{hv}
      NTF optimized for $\sigma=0.2$ & 28.16 & 32.67 & 37.18 \\
      \hlx{hv}
      NTF optimized for $\sigma=0.6$ & 28.26 \Checkmark & 33.01 \Checkmark & 
      38.21 \Checkmark \\
      \hlx{h}
    \end{tabular}
  \end{center}
  \label{tbl:perf-td}
\end{table}%
This new data is illustrated in Tbl~\ref{tbl:perf-td} and is in good accordance
to the previous one. Variations are typically within \unit[1-1.5]{dB}. This
indicates that the nonlinear effects in the modulator should not be ignored,
though. Altogether, we have a confirmation that it is advantageous to design
the \ac{NTF} taking into account the motor model, even if the more realistic
simulations cause the gap over a conventional \ac{NTF} to shrink to
approximately \unit[2.5]{dB}. This is still a \unit[44]{\%} reduction in
noise. Furthermore, the new test reveals that the data fluctuations due to the
nonlinear effects break the already thin advantage related to optimizing the
\ac{NTF} for some specific slip level. Indeed, in the current table, the
optimal values for each column do not lay anymore on the diagonal of rows 2 to
4. This confirms that trying to track a specific slip level with the \ac{NTF}
is a useless over-optimization.

\section{Conclusions}
We have considered the use of ΔΣ modulators in switched mode drives for ac
motors. Specifically, we have looked at how the modulator \ac{NTF} should be
designed to achieve the best possible performance, trying to see if an \ac{NTF}
design practiced taking into account the motor model and the motor loading
condition can be advantageous. From the proposed analysis, it appears that
designing the \ac{NTF} in accordance to the motor model can be convenient,
while trying to have an \ac{NTF} actively tracking the motor slip condition is
probably over-engineering.  From our analysis, designing the \ac{NTF} at a
single, large slip condition is enough to capitalize some non-negligible
advantage (an over \unit[40]{\%} reduction of the quantization noise effect
over the motor winding currents).  Clearly, the proposed analysis is based on
simulation and limited to the parameter set of a specific motor. However, we
expect our conclusions to be valid for many similar setups.




%

\bibliographystyle{IEEEtran}
\bibliography{macros,IEEEabrv,various,sensors,analog,chaos}

\begin{thebibliography}{10}
\providecommand{\url}[1]{#1}
\csname url@samestyle\endcsname
\providecommand{\newblock}{\relax}
\providecommand{\bibinfo}[2]{#2}
\providecommand{\BIBentrySTDinterwordspacing}{\spaceskip=0pt\relax}
\providecommand{\BIBentryALTinterwordstretchfactor}{4}
\providecommand{\BIBentryALTinterwordspacing}{\spaceskip=\fontdimen2\font plus
\BIBentryALTinterwordstretchfactor\fontdimen3\font minus
  \fontdimen4\font\relax}
\providecommand{\BIBforeignlanguage}[2]{{%
\expandafter\ifx\csname l@#1\endcsname\relax
\typeout{** WARNING: IEEEtran.bst: No hyphenation pattern has been}%
\typeout{** loaded for the language `#1'. Using the pattern for}%
\typeout{** the default language instead.}%
\else
\language=\csname l@#1\endcsname
\fi
#2}}
\providecommand{\BIBdecl}{\relax}
\BIBdecl

\bibitem{Holmes:PWMPC-2003}
D.~G. Holmes and T.~A. Lipo, \emph{Pulse Width Modulation for Power
  Converters}.\hskip 1em plus 0.5em minus 0.4em\relax IEEE Press, 2003.

\bibitem{Jacob:EPA-5-7}
B.~Jacob and M.~R. Baiju, ``Spread spectrum modulation scheme for two-level
  inverter using vector quantised space vector-based pulse density
  modulation,'' \emph{IET Electrical Power Application}, vol.~5, no.~7, pp.
  589--596, 2011.

\bibitem{Boys:EPA_B-137-4}
J.~T. Boys and P.~G. Handley, ``Harmonic analysis of space vector modulated
  {PWM} waveforms,'' \emph{Electric Power Applications, IEE Proceedings B.},
  vol. 137, no.~4, pp. 197--204, 1990.

\bibitem{Lowery:TIA-30-2}
T.~F. Lowery and D.~W. Petro, ``Application considerations for {PWM}
  inverter-fed low-voltage induction motors,'' \emph{{IEEE} Trans. Ind. Appl.},
  vol.~30, no.~2, pp. 286--293, 1994.

\bibitem{Stankovic:I3P-90-5}
A.~M. Stankoviç and H.~Lev-Ari, ``Randomized modulation in power electronic
  converter,'' \emph{Proceedings of the IEEE}, vol.~90, no.~5, pp. 782--799,
  May 2002.

\bibitem{Balestra:TIEICEE-E87C-1}
M.~Balestra, A.~Bellini, C.~Callegari, R.~Rovatti, and G.~Setti, ``Chaos-based
  generation of {PWM}-like signals for low-{EMI} induction motor drives:
  Analysis and experimental results,'' \emph{IEICE Transactions on
  Electronics}, vol. E87-C, no.~1, pp. 66--75, Jan. 2004.

\bibitem{Callegari:EL-38-12}
S.~Callegari, R.~Rovatti, and G.~Setti, ``Chaotic modulations can outperform
  random ones in {EMI} reduction tasks,'' \emph{Electronics Letters}, vol.~38,
  no.~12, pp. 543--544, Jun. 2002.

\bibitem{Schreier:UDSDC-2004}
R.~Schreier and G.~C. Temes, \emph{Understanding Delta-Sigma Data
  Converters}.\hskip 1em plus 0.5em minus 0.4em\relax Wiley-IEEE Press, 2004.

\bibitem{Luckjiff:TCAS1-50-8}
G.~Luckjiff and I.~Dobson, ``Hexagonal {S}igma–{D}elta modulation,''
  \emph{{IEEE} Trans. Circuits Syst. {I}}, vol.~50, no.~8, Aug. 2003.

\bibitem{Callegari:TCAS1-Submitted-2012}
S.~Callegari and F.~Bizzarri, ``Output filter aware optimization of the noise
  shaping properties of ΔΣ modulators via semi-definite programming,''
  \emph{{IEEE} Trans. Circuits Syst. {I} \emph{submitted}}, 2012.

\bibitem{Callegari:DeltaSigmaRep-2012}
S.~Callegari, ``An alternative strategy for optimizing the noise transfer
  function of ΔΣ modulators,'' ARCES, University of Bologna, Internal Report,
  2012.

\bibitem{Fitzgerald:EM-1990}
A.~E. Fitzgerald, C.~Kingsley, Jr., and A.~Kusko, \emph{Electrical
  Machinery}.\hskip 1em plus 0.5em minus 0.4em\relax McGraw Hill, 1990.

\bibitem{Bizzarri:ISCAS-2012}
F.~Bizzarri, S.~Callegari, and G.~Gruosso, ``Towards a nearly optimal synthesis
  of power bridge commands in the driving of {AC} motors,'' in
  \emph{Proceedings of ISCAS 2012}, Seoul, May 2012.

\bibitem{Callegari:TSP-58-12}
S.~Callegari, F.~Bizzarri, R.~Rovatti, and G.~Setti, ``On the approximate
  solution of a class of large discrete quadratic programming problems by
  {$\Delta\Sigma$} modulation: the case of circulant quadratic forms,''
  \emph{{IEEE} Trans. Signal Process.}, vol.~58, no.~12, pp. 6126--6139, Dec.
  2010.

\bibitem{Schreier:DELSIG}
\BIBentryALTinterwordspacing
R.~Schreier, \emph{The Delta-Sigma Toolbox}, Analog Devices, 2011, release 7.4,
  also known as ``{DELSIG}''. [Online]. Available:
  \url{http://www.mathworks.com/matlabcentral/fileexchange/}
\BIBentrySTDinterwordspacing

\end{thebibliography}
\end{document}